\documentclass[conference]{IEEEtran}
\IEEEoverridecommandlockouts
\usepackage{cite}
\usepackage{amsmath,amssymb,amsfonts}
\usepackage{algorithmic}
\usepackage{graphicx}
\usepackage{textcomp}
\usepackage{xcolor}
\usepackage{float}
\usepackage{booktabs}
\usepackage{subfigure}
\usepackage{amsmath}
\usepackage{epstopdf}
\usepackage{epsfig}
\usepackage[ruled]{algorithm2e}
\usepackage{multirow}
\usepackage{hyperref}
\usepackage{xcolor}
\usepackage{colortbl} 
\usepackage{hhline} 
\usepackage{longtable}
\usepackage{lscape}
\usepackage{float}
\usepackage{mathtools}
\hypersetup{hidelinks}

\def\BibTeX{{\rm B\kern-.05em{\sc i\kern-.025em b}\kern-.08em
		T\kern-.1667em\lower.7ex\hbox{E}\kern-.125emX}}
\begin{document}

\title{Handover-Aware URLLC UAV Trajectory Planning: A Continuous-Time Trajectory Optimization via Graphs of Convex Sets}

	\author{ Yuqi Ping$^*$, Tingting Zhang$^*$,Tianhao Liang$^*$\\
	$^*$Guangdong Provincial Key Laboratory of Space-Aerial Networking and Intelligent Sensing,\\
	Harbin Institute of Technology, Shenzhen, P. R. China\\
	Email: pingyq@stu.hit.edu.cn, zhangtt@hit.edu.cn, liangth@hit.edu.cn}

\maketitle

\begin{abstract}
	In this paper, we study a cellular-connected unmanned aerial vehicle (UAV) which aims to fly between two predetermined locations while maintaining ultra-reliable low-latency communications (URLLC) for command-and-control (C2) links with terrestrial base stations (BSs). Long-range flights often trigger frequent inter-cell handovers, which may introduce delays and synchronization overhead. We jointly optimize the continuous trajectory and BS association to minimize handovers, path length, and flying time, subject to communication reliability and kinematic constraints. To address this problem, we reformulate it as an optimization based on the graph of convex sets (GCS). First, the URLLC requirement is translated into spatially feasible regions in the flight plane for each BS. And an intersection graph is constructed including the start and goal points. Each graph node is associated with a smooth and dynamically feasible trajectory segment. The trajectory is parameterized in space by Bézier curves and in time by a monotonic Bézier scaling, together with convex constraints that ensure continuity and enforce speed bounds. Next, we impose unit-flow constraints to enforce a single path, and by coupling the resulting binary edge-selection variables with the convex constraints, we obtain a mixed-integer convex program (MICP). Applying a convex relaxation and rounding to the mixed-integer convex program produces nearly globally optimal routes, and a final refinement yields smooth, dynamically feasible trajectories. Simulations verify that the method preserves URLLC connectivity while achieving a clear trade-off between fewer handovers and flight efficiency.
	
\end{abstract}

\section{Introduction}
With the rapid development of the low-altitude economy, unmanned aerial vehicles (UAVs) have been widely deployed in diverse application scenarios, including data collection \cite{11177503}, forest firefighting \cite{ping2025multimodal}, precision agriculture \cite{10531194}, and emergency communications \cite{liang2023age}. Cellular-connected UAVs are particularly attractive for maintaining reliable command-and-control (C2) links, owing to the wide-area coverage, high-capacity backhaul, and beyond-visual-line-of-sight (BVLOS) operation provided by terrestrial base stations (BSs). However, long-range flights inevitably traverse cells and thus trigger frequent handovers, which may cause excessive latency, synchronization overhead, and ping–pong effects \cite{madelkhanova2022optimization}. Consequently, trajectory designs that explicitly couple mobility control with BS association are required.

To address the above tension, extensive research has been conducted on trajectory optimization for cellular-connected UAVs under communication constraints. A representative line of work reformulates connectivity-constrained path planning as a graph problem. Under a distance-threshold connectivity assumption, \cite{zhang2018cellular} formulated a hybrid framework that integrates convex optimization with graph-based routing and developed an optimal algorithm with non-polynomial time complexity. Building on this graph-based formulation, subsequent investigations have further incorporated communication outages \cite{zhang2019trajectory}, battery limitations \cite{im2025trajectory}, propulsion energy considerations \cite{yang2021efficient},  and the objective of minimizing handovers under a maximum mission-time constraint \cite{du2025handover}. These methods discretize the continuous-time UAV trajectory planning problem into a finite waypoint selection problem, typically assume a constant speed between successive waypoints, and decompose the overall optimization into several subproblems, such as BS selection and waypoint refinement. Although this decomposition enables efficient search over the constructed graph, the resulting paths may violate continuous-time dynamic feasibility, for example, curvature, acceleration, or turn-rate limits, and, due to the discretization and sequential decomposition, the obtained solutions are generally not guaranteed.

Moreover, most existing investigations ensure link quality using average SNR or throughput, whereas the safety-critical C2 requires finite-blocklength ultra-reliable low-latency communications (URLLC), so the achievable rate no longer depends on SNR alone but also on the latency budget and channel dispersion. Ignoring URLLC can produce trajectories that satisfy average-rate or SNR targets but violate packet-level reliability and latency constraints, leading to transient outages during maneuvers or near cell edges that compromise the safety of C2\cite{salehi2022ultra}. While \cite{11174804} investigated safety-aware self-triggered model predictive control for BS--UAV systems under URLLC constraints, BS handovers were not considered.

Motivated by the above issues, we aim to jointly design multiple BS association and continuous UAV trajectory planning under URLLC constraints, so as to guarantee dynamic feasibility while minimizing handovers, path length, and flight time. The main contributions of this work are outlined as follows.
\begin{itemize}
	\item We first introduce a handover-aware URLLC UAV trajectory planning framework that jointly optimizes BS association and continuous-time UAV trajectory under finite-blocklength URLLC constraints. The framework ensures C2-link reliability at all times while enforcing bounded velocity and high-order trajectory curvature smoothness. In our framework, handovers, path length, and flight time can be jointly optimized while maintaining dynamical feasibility.
	\item Subsequently, we formulate a Bézier-parameterized graph-of-convex-sets (GCS) approach \cite{marcucci2023motion}. Unlike traditional waypoint-based optimization methods, it jointly parameterizes the spatial trajectory and time scaling, encodes URLLC-feasible regions as a convex intersection graph, and casts the problem as a mixed-integer convex program with quadratic and second-order cone constraints. Applying a convex relaxation and rounding to the mixed-integer convex program produces nearly globally optimal routes, and a final refinement yields smooth, dynamically feasible trajectories while ensuring computational efficiency.
	\item Numerical simulations are provided to demonstrate that the proposed method generates smooth, physically feasible trajectories that reduce handovers while maintaining high path and time efficiency. In addition, by tuning the cost weights, the method can further optimize time-optimal trajectories under the constraint of minimizing the number of handover, highlighting the method’s generalizability and practical applicability.
\end{itemize}

\section{System Model and Problem Description}
\subsection{General Descriptions}

\begin{figure}[h]
	\centering
	\includegraphics[width=0.45\textwidth]{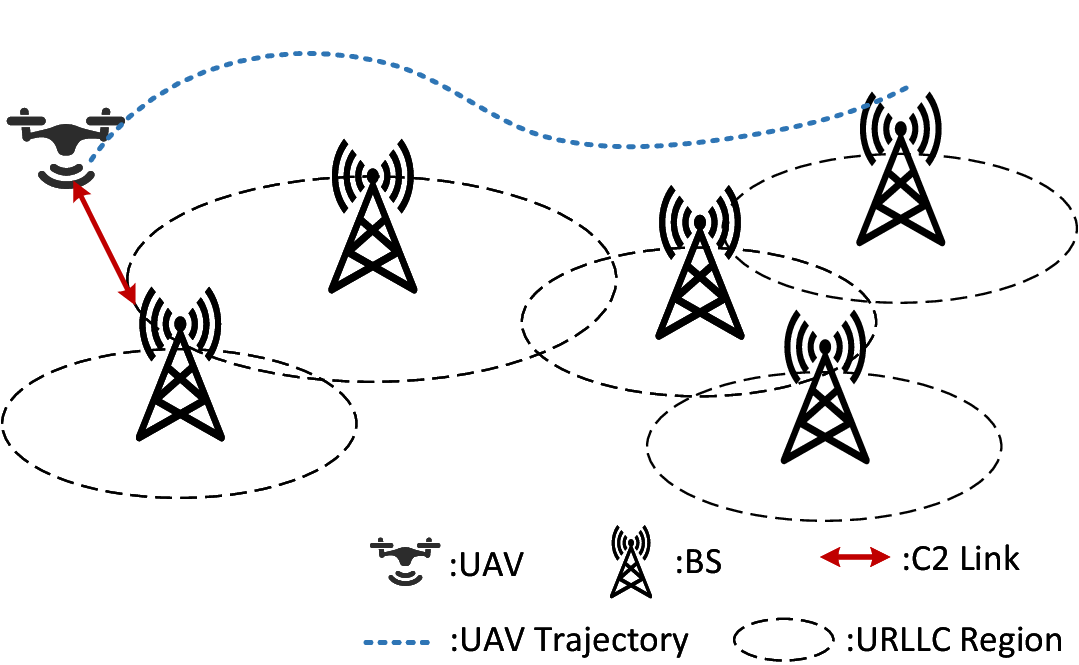}
	\caption{Illustration of the proposed cellular-connected UAV system.}
	\label{fig:system model}
\end{figure}

As illustrated in Fig. \ref{fig:system model}, we consider a handover-aware, URLLC-constrained cellular-connected UAV that travel from a prescribed start point to a target point. The feasible flight airspace is $\mathcal{Q}\subset\mathbb{R}^3$. The UAV departs from $q_s\in\mathbb{R}^3$, flies at a constant altitude $H$, and reaches $q_g\in\mathbb{R}^3$. To ensure safety, the UAV–BS C2 link must satisfy URLLC constraints at all times. Let $\mathcal{B}=\{1,\ldots,M\}$ with $M\ge1$ denote the set of BSs. At any time, the UAV is associated with exactly one BS and may perform handovers among BSs along the trajectory. Accordingly, we formulate a handover-aware planning optimization problem that guarantees continuous URLLC connectivity during the flight from $q_s$ to $q_g$ while jointly minimizing the number of handovers, the path length, and the flight time.

\subsection{UAV Kinematic Model}
We model the UAV as a double integrator. Let $\boldsymbol{q}(t) = [x(t),\, y(t),\, z(t)]^\top \in \mathbb{R}^3$ denote position, 
$\boldsymbol{v}(t) = \dot{\boldsymbol{q}}(t) \in \mathbb{R}^3$ denote velocity, and 
$\boldsymbol{a}(t) = \dot{\boldsymbol{v}}(t) = \ddot{\boldsymbol{q}}(t) \in \mathbb{R}^3$ denote control input. 
The dynamics are
\begin{equation}
	\dot{\boldsymbol{q}}(t) = \boldsymbol{v}(t), 
	\qquad 
	\dot{\boldsymbol{v}}(t) = \boldsymbol{a}(t).
\end{equation}
The UAV is subject to a maximum speed constraint
\begin{equation}
	\|\boldsymbol{v}(t)\| \le v_{\max}, \qquad \forall\,t.
\end{equation}
In our planning problem, the flight altitude is fixed to $z(t)=H$ and the UAV position satisfies $[x(t),y(t),H]^\top\in\mathcal{Q}$.

\subsection{UAV to BS Communication Model}

Following the UAV-to-ground communication model in \cite{al2014optimal}, without loss of generality, we characterize the UAV-BS communication  channel as a combination of line-of-sight (LoS) and non-line-of-sight (NLoS) components. For each BS $i \in \mathcal{B}$, its three-dimensional position is defined as
\begin{equation}
	\boldsymbol{q}_{i} = [x_{i},\,y_{i},\,z_{i}]^\top \in \mathbb{R}^3,
\end{equation}
where $x_i$ and $y_i$ represent the ground-plane coordinates and $z_i$ denotes the antenna height of BS $i$.

At time SLOT $t$, the 3D link distance and elevation angle between the UAV and BS $i$ are given by
\begin{align}
	d_{i}(t) &= \big\|\,\boldsymbol{q}(t)-\boldsymbol{q}_{i}\,\big\|,\\
	\theta_{i}(t) &= \arctan\!\Big(\frac{z(t)-z_{i}}{\sqrt{(x(t)-x_{i})^2+(y(t)-y_{i})^2}}\Big).
\end{align}

To capture the effect of the elevation angle on LoS occurrence, we adopt a widely used empirical model in which the LoS probability follows a logistic function of the elevation angle. Specifically, the LoS and NLoS probabilities are modeled as
\begin{align}
	p^{\mathrm{LoS}}_i(t) &= \frac{1}{1 + a\,\exp\!\big(-b(\theta_i(t)-a)\big)},\\
	p^{\mathrm{NLoS}}_i(t) &= 1 - p^{\mathrm{LoS}}_i(t),
\end{align}
where $a$ and $b$ are environment-dependent parameters determined by the type of terrain.

Accordingly, the average path loss between the UAV and BS $i$ can be expressed as
\begin{equation}
	L_{i}(t) = p^{\mathrm{LoS}}_i(t)\,L_{i}^{\mathrm{LoS}}(t)
	+ p^{\mathrm{NLoS}}_i(t)\,L_{i}^{\mathrm{NLoS}}(t),
\end{equation}
where the LoS and NLoS components are respectively given by
\begin{align}
	L_{i}^{\mathrm{LoS}}(t)   &= \Big(\tfrac{4\pi f_c\, d_{i}(t)}{c}\Big)^2 \eta^{\mathrm{LoS}},\\
	L_{i}^{\mathrm{NLoS}}(t) &= \Big(\tfrac{4\pi f_c\, d_{i}(t)}{c}\Big)^2 \eta^{\mathrm{NLoS}},
\end{align}
with $f_c$ denoting the carrier frequency, $c$ the speed of light, and $\eta^{\mathrm{LoS}}$, $\eta^{\mathrm{NLoS}}$ the excessive path loss factors corresponding to the LoS and NLoS cases, respectively.

Given the BS transmit power $P_i$, receiver antenna gain $g_r$, and receiver noise power $\sigma^2$, the instantaneous received signal-to-noise ratio (SNR) at BS $i$ can be calculated as
\begin{equation}
	\gamma_i(\boldsymbol{q}) = \frac{g_r\,P_{i}}{L_{i}\,\sigma^2}.
\end{equation}

\subsection{Short-Packet URLLC Constraint}

We impose a URLLC requirement on the UAV--BS C2 link in terms of latency, reliability, and rate. 
Let $L$ denote the end-to-end latency budget and $\varepsilon_{\max}$ the target decoding error probability. 
Given system bandwidth $B$ and transmission duration $\tau$, the finite-blocklength is
\begin{equation}
	n = B\tau, \qquad 0< \tau \le L, \qquad 0< \varepsilon \le \varepsilon_{\max},
	\label{eq:blocklength}
\end{equation}
where $\varepsilon$ is the packet  error probability, the reliability constraint enforces $\varepsilon\le\varepsilon_{\max}$, while the latency budget enforces $\tau\le L$.

Under the normal approximation for finite blocklength coding \cite{durisi2016toward}, the achievable per-channel-use rate is
\begin{equation}
	R(\gamma,n,\varepsilon)
	\approx \log_2(1+\gamma)\;-\;\sqrt{\frac{V(\gamma)}{n}}\,Q^{-1}(\varepsilon)\;+\;\frac{\log_2 n}{2n},
	\label{eq:Rfb}
\end{equation}
where the channel dispersion is
\begin{equation}
	V(\gamma) = \frac{\gamma(\gamma+2)}{(1+\gamma)^2}\,(\log_2 e)^2,
\end{equation}
$\gamma$ is the receive SNR, and $Q^{-1}(\cdot)$ is the inverse Gaussian $Q$-function.

Let $R_{\mathrm{req}}$ denote the required coding rate for the C2 link. 
The URLLC feasibility condition is
\begin{equation}
	R(\gamma,n,\varepsilon_{\max}) \;\ge\; R_{\mathrm{req}},
	\label{eq:urlcc-feasible}
\end{equation}
For a given $(n,\varepsilon_{\max})$, define the minimum required SNR as
\begin{equation}
	\gamma_{\min}(n) \;\triangleq\; \inf\left\{\gamma \ge 0:\;
	R(\gamma,n,\varepsilon_{\max}) \ge R_{\mathrm{req}}\right\}.
\end{equation}

Let $\gamma_i(\boldsymbol{q})$ be the received SNR from BS $i$ at position $\boldsymbol{q}$ under the air-to-ground channel model. 
Each BS induces a URLLC-feasible region
\begin{equation}
	\mathcal{C}_i \;\triangleq\; \big\{\, \boldsymbol{q} \in \mathcal{Q}:\; \gamma_i(\boldsymbol{q}) \ge \gamma_{\min} \,\big\},
\end{equation}
and the UAV trajectory must satisfy $\boldsymbol{q}(t)\in\bigcup_{b\in\mathcal{B}}\mathcal{C}_b$ to maintain short-packet URLLC connectivity under the blocklength, latency, and reliability constraints.

\subsection{Problem Description}

We jointly optimize the UAV trajectory and the BS association to balance the number of handovers, the total flight time, and the path length. 
Let $b(t)\in\mathcal{B}$ denote the serving BS index at time $t$, where $b(\cdot)$ is piecewise-constant and right-continuous. 
$T$ denotes the total flight time of the UAV. 
Define the handover time set
\begin{equation}
	\mathcal{T}_{\mathrm{ho}}\triangleq\{\,t\in(0,T]:\, b(t^-)\neq b(t)\,\},
\end{equation}
and let $N\triangleq|\mathcal{T}_{\mathrm{ho}}|$ denote the total number of handovers.
To quantify the geometric path cost while keeping later relaxations convex, we adopt the path-length functional
\begin{equation}
\mathcal{L}(\boldsymbol{q}) \triangleq \int_{0}^{T} \|\boldsymbol{v}(t)\|_2\,\mathrm{d}t .
\end{equation}
With nonnegative weights $(\lambda_{\mathrm{ho}},\beta,\alpha)$ trading off handover count, total time, and path length, the handover-aware URLLC-constrained trajectory optimization is
\begin{subequations}\label{prob:P1}
	\begin{align}
		\text{(P1)}\quad 
		&\min_{\substack{\boldsymbol{q},\,\boldsymbol{v},\,\boldsymbol{a},\\ b,\,T}}
		\ \ \lambda_{\mathrm{ho}}\,N\ +\ \beta\,T\ +\ \alpha\,\mathcal{L}(\boldsymbol{q}) \\[2pt]
		\text{s.t.}\quad 
		& \dot{\boldsymbol{q}}(t)=\boldsymbol{v}(t),\quad \dot{\boldsymbol{v}}(t)=\boldsymbol{a}(t), \hfill \forall t, \label{eq:dyn}\\
		& \|\boldsymbol{v}(t)\|_2\le v_{\max}, \hfill \forall t, \label{eq:speed}\\
		& \boldsymbol{q}(0)=\boldsymbol{q}_s,\ \boldsymbol{q}(T)=\boldsymbol{q}_g, \label{eq:bc}\\
		& z(t)=H,\ \boldsymbol{q}(t)\in\mathcal{Q}, \hfill \forall t, \label{eq:feasible}\\
		& b(t)\in\mathcal{B}, \hfill \forall t, \label{eq:assoc}\\
		& \gamma_{\,b(t)}\big(\boldsymbol{q}(t)\big)\ \ge\ \gamma_{\min}, \hfill \forall t. \label{eq:urlcc}
	\end{align}
\end{subequations}
Here, $\gamma_{\,b(t)}(\boldsymbol{q}(t))$ denotes the received SNR from the serving BS at the UAV's position at time $t$. P1 is thus a mixed discrete–continuous optimization problem that jointly determines the discrete BS association and the continuous UAV trajectory. Unlike waypoint discretization and graph-based shortest paths, we develop a GCS-based optimization algorithm that searches directly over smooth Bézier-spline trajectories in continuous space, enforces the URLLC constraints for all times, and exploits a tight convex relaxation to obtain certificates of near-global optimality with scalability that grows only polynomially with trajectory resolution.

\section{GCS-Based Reformulation with Bézier-Parameterized Trajectory}
\label{sec:GCS_Bezier}
In this section, we reformulate P1 as a graph-of-convex-sets program. We lift the feasible flight plane into an intersection graph in which each vertex represents B\'ezier-parameterized trajectory segments under convex constraints, while each directed edge encodes inter-segment continuity and handover-aware costs.
\subsection{Convex Intersection Graph of URLLC-Feasible Regions}
\label{subsec:graph}

On the constant-altitude plane $z=H$, positions satisfying the URLLC SNR threshold $\gamma_{\min}$ for BS $i$ obey
\begin{equation}
	L_i \ \le\ \frac{g_r P_i}{\sigma^2\,\gamma_{\min}}.
\end{equation}
Since $L_i$ increases monotonically with horizontal separation when $z=H$ is fixed, there exists a maximal horizontal radius $\rho_i$ such that all points within $\rho_i$ of the BS ground projection meet the inequality above. Let $\boldsymbol{c}_i \!\triangleq\! (x_i,y_i)$ denote the ground-plane coordinates of BS $i$ and define
\begin{equation}
	\rho_i\ \triangleq\ \sup\Big\{\,r\ge0:\ L_i\big(\sqrt{r^2+(H-z_i)^2}\big)\ \le\ \tfrac{g_r P_i}{\sigma^2\,\gamma_{\min}}\,\Big\}.
	\label{eq:rho_def}
\end{equation}
The URLLC-feasible set induced by BS $i$ on $z=H$ is then the closed disk
\begin{equation}
	\mathcal{R}_i \ \triangleq\ \big\{(x,y,H)\in\mathcal{Q}:\ \|(x,y)-\boldsymbol{c}_i\|_2 \le \rho_i\big\},
	\label{eq:disk_def}
\end{equation}
and the overall feasible region is the union $\bigcup_{i\in\mathcal{B}}\mathcal{R}_i$.

We capture overlaps among these convex sets via a directed intersection graph
\begin{equation}
	\mathcal{G}=(\mathcal{V}\cup\{s,g\},\mathcal{E}),\qquad
	\mathcal{V}=\{\,i\in\mathcal{B}:\ \mathcal{R}_i\neq\emptyset\,\}.
\end{equation}
Here, $s$ and $g$ are nodes corresponding to $\boldsymbol{q}_s$ and $\boldsymbol{q}_g$. Edges are created by nonempty overlaps and by start target inclusion
\begin{align}
	&(i,j)\in\mathcal{E}\ \iff\ \mathcal{R}_i\cap\mathcal{R}_j\neq\emptyset,\ \ i\neq j,\nonumber\\
	&(s,i)\in\mathcal{E}\ \iff\ \boldsymbol{q}_s\in\mathcal{R}_i,\nonumber\\
	&(i,g)\in\mathcal{E}\ \iff\ \boldsymbol{q}_g\in\mathcal{R}_i.
	\label{eq:IG_edges}
\end{align}
Any $s$–$g$ path in $\mathcal{G}$ yields a sequence of URLLC-feasible sets through which the UAV can travel while maintaining $\boldsymbol{q}\in\mathcal{Q}$, the number of internal edges on such a path equals the number of handovers. This intersection-graph representation underpins the subsequent GCS-based trajectory and handover optimization.

\subsection{Node-Level Trajectory Representation and Convex Kinematic Constraints}
\label{subsec:node}
We parameterize both the shape curve and the time-scaling curve with Bézier curves, enabling convex and exact enforcement of safety and kinematic constraints over continuous time. Each node $i\in\mathcal V$ carries two degree-$m$ Bézier parameterizations on $\xi\in[0,1]$: a planar shape curve and a strictly increasing time-scaling curve
\begin{align}
	\boldsymbol r_i(\xi)&=\sum_{k=0}^{m}\boldsymbol r_{i,k}\,B_k^{(m)}(\xi), \quad \boldsymbol r_{i,k}\in\mathbb R^2,\\
	h_i(\xi)&=\sum_{k=0}^{m} h_{i,k}\,B_k^{(m)}(\xi),\qquad h_{i,k}\in\mathbb R,\\
	B_k^{(m)}(\xi)&=\binom{m}{k}\xi^{k}(1-\xi)^{m-k}.
\end{align}
The physical segment is $\boldsymbol q_i(t)=[\boldsymbol r_i(\xi)^\top,H]^\top$ with the time reparameterization $t=h_i(\xi)$.

For any derivative order $p=0,1,\ldots,m$, both $\boldsymbol r_i^{(p)}(\xi)$ and $h_i^{(p)}(\xi)$ are Bézier curves of degree $m-p$, whose control points are linear combinations of the original control points. Define the control points of the $p$-th derivative Bézier curves by
\begin{align}
	\boldsymbol r^{(p)}_{i,k} \triangleq \frac{m!}{(m-p)!}\sum_{j=0}^{p}(-1)^{p-j}\binom{p}{j}\,\boldsymbol r_{i,k+j}, &\\
	h^{(p)}_{i,k} \triangleq \frac{m!}{(m-p)!}\sum_{j=0}^{p}(-1)^{p-j}\binom{p}{j}\, h_{i,k+j},&
\end{align}
where $	k=0,\ldots,m-p,$ so that
\begin{align}
	\boldsymbol r_i^{(p)}(\xi)=\sum_{k=0}^{m-p}\boldsymbol r^{(p)}_{i,k}\,B_k^{(m-p)}(\xi),\\
	h_i^{(p)}(\xi)=\sum_{k=0}^{m-p} h^{(p)}_{i,k}\,B_k^{(m-p)}(\xi).
\end{align}

By the convex-hull property of Bézier curves, which follows from the nonnegativity and partition-of-unity of the Bernstein basis, if all shape control points satisfy $\boldsymbol r_{i,k}\in\mathcal R_i$ for $k=0,\ldots,m$, then the entire Bézier segment lies in $\mathcal R_i$. Strict time monotonicity is ensured by requiring the first derivative control points of the time-scaling curve to be positive, $h^{(1)}_{i,k}>0$ for $k=0,\ldots,m-1$, which makes $h_i$ strictly increasing and invertible. The maximum-speed constraint is enforced via second-order cone inequalities, that is
\begin{equation}
	\|\boldsymbol r^{(1)}_{i,k}\|_2 \le v_{\max}\, h^{(1)}_{i,k}, \qquad k=0,\ldots,m-1,
\end{equation}
which ensures that $\|\dot{\boldsymbol q}_i(t)\|\le v_{\max}$ over the entire segment.

\subsection{Edge Coupling and Handover-Aware Edge Lengths}
\label{subsec:edge}
Edges encode boundary conditions, inter-segment continuity, and convex costs. For any edge $(s,i)\in\mathcal E$ outgoing from the source, fix the initial waypoint, initial time, and zero initial velocity by
\begin{equation}
	\boldsymbol r_{i,0}=\boldsymbol q_s,\qquad h_{i,0}=0,\qquad \boldsymbol r^{(1)}_{i,0}=\boldsymbol 0.
\end{equation}
For any edge $(i,g)\in\mathcal E$ incoming to the sink, fix the terminal waypoint and zero terminal velocity by
\begin{equation}
	\boldsymbol r_{i,m}=\boldsymbol q_g,\qquad \boldsymbol r^{(1)}_{i,m-1}=\boldsymbol 0.
\end{equation}
On any internal edge $(i,j)\in\mathcal E$ with $i,j\in\mathcal V$, impose $C^\eta$ inter-segment continuity by matching endpoint derivative control points up to order $\eta$
\begin{equation}
	\boldsymbol r^{(p)}_{i,\,m-p}=\boldsymbol r^{(p)}_{j,\,0},\ 
	h^{(p)}_{i,\,m-p}=h^{(p)}_{j,\,0},\  p=0,1,\ldots,\eta.
\end{equation}

To align with the continuous-time tradeoff, assign a convex length to each edge. Leaving the source incurs no cost, that is
\begin{equation}
	\ell_{(s,i)}=0.
\end{equation}
For any internal edge $(i,j)$, the length aggregates geometric effort, elapsed time, a unit handover penalty, and a second-order smoothing regularizer
\begin{align}
	\ell_{(i,j)}\ &=\
	\alpha \sum_{k=0}^{m-1}\big\|\boldsymbol r^{(1)}_{i,k}\big\|_2^{2}
	\ +\ \beta\,\big(h_{i,m}-h_{i,0}\big)\ +\ \lambda_{\mathrm{ho}} \nonumber\\
	&\ +\ \gamma_{\mathrm{sm}}\!\left(\sum_{k=0}^{m-2}\big\|\boldsymbol r^{(2)}_{i,k}\big\|_2^{2}
	\ +\ \sum_{k=0}^{m-2}\big(h^{(2)}_{i,k}\big)^{2}\right)\!,
\end{align}
where the second-order smoothing regularizer penalizes curvature variations and promotes overall trajectory smoothness. $\alpha,\beta,\lambda_{\mathrm{ho}}\ge0$ weight geometric effort, time, and handovers, and $\gamma_{\mathrm{sm}}\ge0$ sets the strength of this regularization. Going from an internal node directly to the sink carries the same geometric, temporal, and smoothing terms but no handover penalty
\begin{align}
	\ell_{(i,g)}\ &=\
	\alpha \sum_{k=0}^{m-1}\big\|\boldsymbol r^{(1)}_{i,k}\big\|_2^{2}
	\ +\ \beta\,\big(h_{i,m}-h_{i,0}\big)\nonumber\\
	&\ +\ \gamma_{\mathrm{sm}}\!\left(\sum_{k=0}^{m-2}\big\|\boldsymbol r^{(2)}_{i,k}\big\|_2^{2}
	\ +\ \sum_{k=0}^{m-2}\big(h^{(2)}_{i,k}\big)^{2}\right).
\end{align}
All terms above are linear or quadratic in the control points.
\subsection{Flow-Coupled GCS Program, Convex Relaxation, and Path Extraction}
\label{subsec:micp_relax}

To couple discrete path selection with continuous trajectory optimization, we introduce binary flow variables $y_{ij}\in\{0,1\}$ for each directed edge $(i,j)\in\mathcal E$ in the intersection graph $\mathcal G=(\mathcal V\cup\{s,g\},\mathcal E)$, and enforce a unit $s$–$g$ flow
\begin{subequations}\label{eq:unitflow}
	\begin{align}
		&\sum_{(s,j)\in\mathcal E}y_{sj}=1,\qquad \sum_{(i,g)\in\mathcal E}y_{ig}=1,\\
		&\sum_{(v,j)\in\mathcal E}y_{vj}-\sum_{(i,v)\in\mathcal E}y_{iv}=0,\quad \forall v\in\mathcal V,\\
		&y_{ij}\in\{0,1\},\quad \forall (i,j)\in\mathcal E.
	\end{align}
\end{subequations}

Each node $i\in\mathcal V$ maintains the Bézier control points $\{\boldsymbol r_{i,k}\}_{k=0}^{m}$ and $\{h_{i,k}\}_{k=0}^{m}$ for the shape and time-scaling curves, subject to convex set membership, strict time monotonicity, and second-order cone speed constraints, as previously defined. Internal edges $(i,j)\in\mathcal E$ enforce $C^\eta$ continuity between neighboring segments, while source and sink edges fix endpoint boundary conditions. The geometric, temporal, and handover costs are combined into the convex edge length $\ell_{(i,j)}$, and the overall optimization problem becomes a mixed-integer convex program
\begin{subequations}\label{prob:GCS-MICP}
	\begin{align}
		\text{(P2)}\quad&\min_{\{y_{ij}\},\,\{\boldsymbol r_{i,k}\},\,\{h_{i,k}\}}\quad \sum_{(i,j)\in\mathcal E}\ \ell_{(i,j)}\,y_{ij}\\[2pt]
		\text{s.t.}\quad &\text{flow constraints (\ref{eq:unitflow})},\\
		&\boldsymbol r_{i,k}\in\mathcal R_i,h^{(1)}_{i,k}\ge 0,k=0,\ldots,m-1,\\
		&\big\|\boldsymbol r^{(1)}_{i,k}\big\|_2\le v_{\max}h^{(1)}_{i,k},k=0,\ldots,m-1,\\
		&\boldsymbol r^{(p)}_{i,m-p}=\boldsymbol r^{(p)}_{j,0}, h^{(p)}_{i,m-p}=h^{(p)}_{j,0}, p=0,\ldots,\eta,\\
		&\boldsymbol r_{i,0}=\boldsymbol q_s,\ h_{i,0}=0,\ \boldsymbol r^{(1)}_{i,0}=\boldsymbol 0,\\
		&\boldsymbol r_{i,m}=\boldsymbol q_g,\ \boldsymbol r^{(1)}_{i,m-1}=\boldsymbol 0.
	\end{align}
\end{subequations}

P2 is posed as a mixed-integer convex program (MICP). When the binary edge-selection variables are relaxed to $y_{ij}\!\in[0,1]$, the squared-norm and speed constraints admit a convex second-order cone (SOCP) formulation, yielding a computable lower bound on the optimal cost. Empirically, solutions to this relaxation concentrate most of the flow on a small set of edges, which often approximates a single discrete path \cite{marcucci2024shortest}. We then extract an $s$ to $g$ path via randomized depth-first rounding guided by the outgoing fractional flows at each node. After a path is obtained, we prune unused nodes and edges and solve a final convex refinement problem along the selected route to adjust the Bézier control points. In this stage we minimize the sum of active edge costs subject to all convex constraints.

We implement the optimization using Drake’s Graph-of-Convex-Sets (GCS) framework \cite{marcucci2023motion} with MOSEK as the solver backend, which efficiently handles second-order cone and quadratic constraints. This combination exhibits good computational performance and scalability.

\begin{table}[h]
	\centering
	\caption{Simulation parameters}
	\begin{tabular}{lc}
		\hline
		\textbf{Parameter} & \textbf{Value} \\
		\hline
		Carrier frequency & $f_c = 3.3~\text{GHz}$ \\
		BS transmit power & $P_i = 0.09~\text{W}$  \\
		Receiver gain & $g_r = 1$ \\
		Noise power & $\sigma^2=7.21 \times 10^{-16}~\text{W}$ \\
		Bandwidth & $B = 180~\text{kHz}$ \\
		Transmission duration & $\tau = 1~\text{ms}$ \\
		Blocklength & $n = 180$ \\
		URLLC error probability & $\varepsilon_{\max} = 10^{-5}$ \\
		Required rate & $R_{\mathrm{req}} = 0.5~\text{bit/s/Hz}$ \\
		LoS probability parameter & $a = 12.08,b = 0.11$ \\
		LoS excess loss & $\eta^{\mathrm{LoS}} = 3~\text{dB}$ \\
		NLoS excess loss & $\eta^{\mathrm{NLoS}} = 25~\text{dB}$ \\
		\hline
	\end{tabular}
	\label{tab:sim_params_urlcc}
\end{table}

\section{Numerical Results}
In this section,we consider a cellular-connected UAV scenario within a $5~\text{km}\times5~\text{km}$ square region. 
A total of $M=30$ BSs are randomly deployed on the ground plane over this area, and each BS is assigned an antenna height uniformly within $[0,200]~\text{m}$. The UAV flies at a fixed altitude $H=300~\text{m}$, starting from the initial position $\boldsymbol{q}_s=[250,\,2250,\,300]^\top~\text{m}$ and traveling toward the target position $\boldsymbol{q}_g=[4000,\,4000,\,300]^\top~\text{m}$. The other main simulation parameters are summarized in Table \ref{tab:sim_params_urlcc}.

\begin{figure}[h]
	\centering
	\includegraphics[trim=130 0 140 20, clip, width=0.4\textwidth]{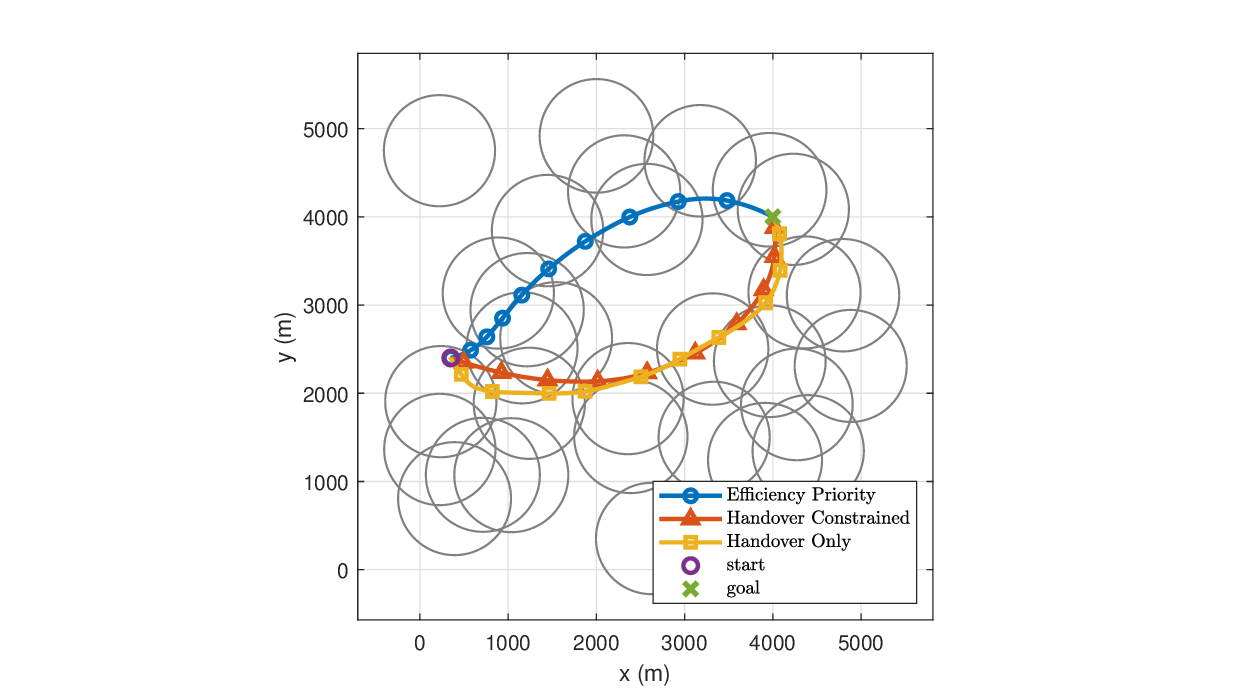}
	\caption{Optimized UAV trajectories under three weighting configurations: Efficiency-Priority, Handover-Constrained, and Handover-Only within a 5 km × 5 km cellular layout.}
	\label{fig:trajectory_comparison}
\end{figure}
\vspace{-10pt}

We compare three weight settings for $(\alpha,\beta,\lambda_{\mathrm{ho}},\gamma_{\mathrm{sm}})$ that is Efficiency Priority, Handover Constrained, and Handover Only. All produce smooth and continuous trajectories, as shown in Fig.~\ref{fig:trajectory_comparison}. Efficiency Priority $(0.5,1,0.1,0.005)$ yields $447.68~\text{s}$ flight time, $4304.42~\text{m}$ length, and $7$ handovers, offering a balanced compromise. Handover Constrained $(0.5,1,10000,0.005)$ achieves the minimal $6$ handovers and, among feasible $6$-handover solutions, yields a comparatively short path $4774.43\,\mathrm{m}$ and comparatively short flight time $500.56\,\mathrm{s}$. Handover Only $(0,0,10000,0.005)$ also enforces $6$ handovers by lingering near cell-center regions, lacking distance and time penalties, it incurs much larger time and distance. 

\begin{figure}[h]
	\centering
	\includegraphics[trim=40 0 50 10, clip,width=0.45\textwidth]{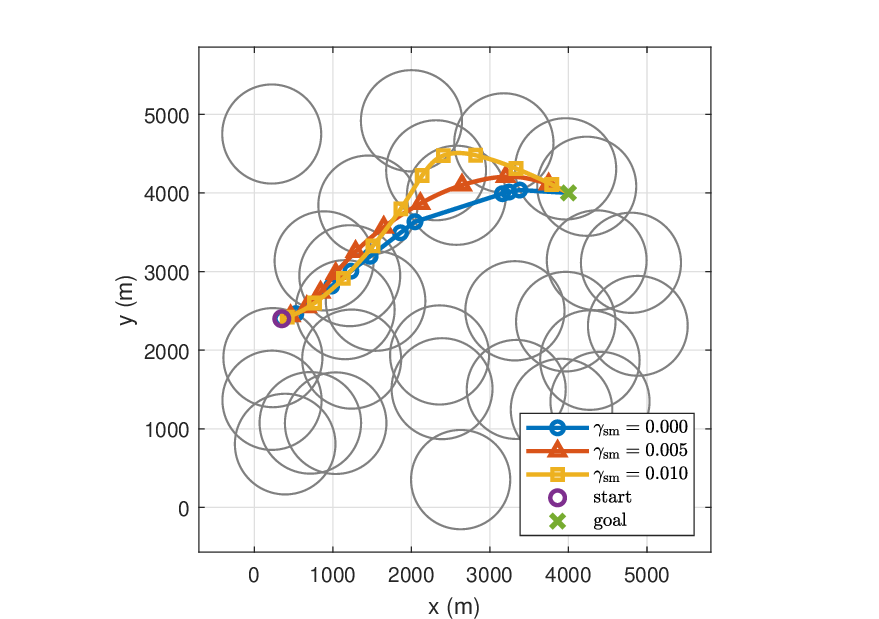}
	\caption{Trajectory comparison under different smoothing regularization strengths.}
	\label{fig:trajectory_sm}
\end{figure}

\begin{figure}[h]
	\centering
	\includegraphics[trim=25 0 30 20, clip,width=0.4\textwidth]{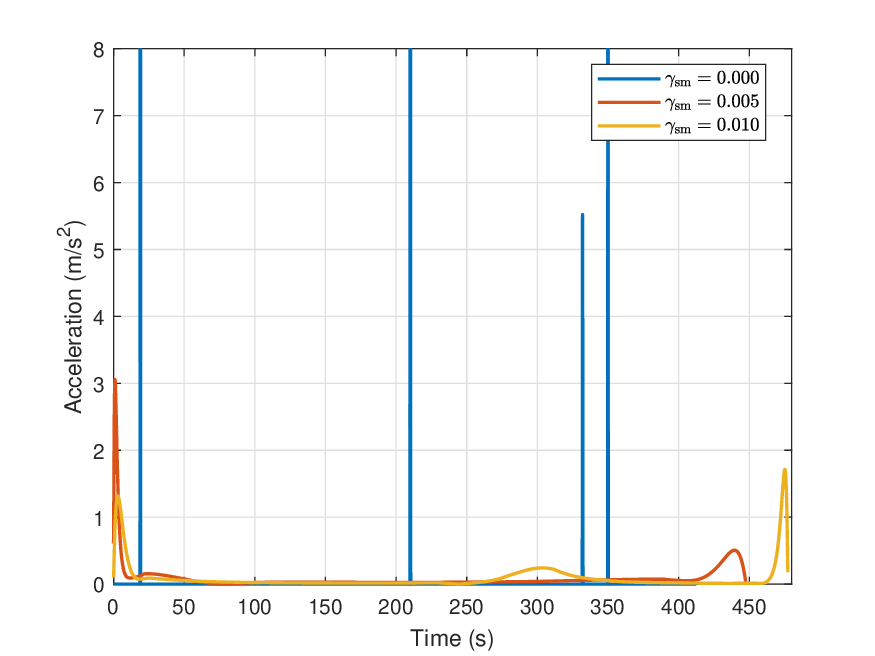}
	\caption{Acceleration profiles corresponding to different smoothing regularization strengths.}
	\label{fig:acc_sm}
\end{figure}

To further examine the effect of smoothing, we fix $(\alpha,\beta,\lambda_{\mathrm{ho}})=(0.5,1,0.1)$ and vary the regularization strength $\gamma_{\mathrm{sm}}\in\{0.000,\,0.005,\,0.010\}$. The resulting trajectories are shown in Fig.~\ref{fig:trajectory_sm}, and the acceleration time histories are presented in Fig.~\ref{fig:acc_sm}. Increasing $\gamma_{\mathrm{sm}}$ systematically enhances geometric smoothness and suppresses peak accelerations, thereby reducing the risk of dynamic infeasibility due to excessive control effort. The accompanying trade-off is that stronger smoothing generally increases total path length and flight time. In our experiments, $\gamma_{\mathrm{sm}}=0$ produces the sharpest turns and the largest acceleration spikes, $\gamma_{\mathrm{sm}}=0.005$ yields a favorable balance, and $\gamma_{\mathrm{sm}}=0.010$ provides the strongest regularization at the cost of the largest increases in distance and duration.

All the above optimization scenarios can be solved within 15~s on a PC equipped with an Intel~i7-13700K processor and~32~GB of~DDR5-6800~MHz memory.

\vspace{-20pt}

\section{Conclusion}
In this paper, we proposed a novel GCS optimization framework to jointly design the continuous trajectory and BS association for a cellular-connected UAV under stringent finite-blocklength URLLC constraints. By translating URLLC requirements into convex feasible regions and parameterizing the trajectory and time-scaling using Bézier curves, we formulated the task of minimizing handovers, flight time, and path length as a tractable MICP with second-order cone constraints and linear constraints. Numerical simulations demonstrated that our method efficiently generates smooth, dynamically feasible trajectories that guarantee continuous URLLC connectivity, while clearly exposing the trade-offs between handover frequency and flight efficiency. Future work will extend the proposed approach to non-isotropic, time-varying wireless channels.

	\bibliographystyle{IEEEtran}
	\bibliography{reference}
	
\end{document}